\documentclass[twocolumn,superscriptaddress,aps,prb,]{revtex4-1}
\usepackage{bm}
\usepackage{amssymb,amsmath}
\usepackage{comment}
\usepackage{times}
\usepackage{graphicx}
\usepackage{array}

\begin{document}

\title{Effect of surface disorder on the chiral surface states of a three-dimensional quantum Hall system}

\author{Chao Zheng}
\email{zhengchaonju@gmail.com}
\affiliation{Zhejiang Institute of Modern Physics, Zhejiang University, Hangzhou 310027, China}

\author{Kun Yang}
\affiliation{Physics Department and National High Magnetic Field Laboratory, Florida State University, Tallahassee, Florida 32306, USA}

\author{Xin Wan}
\affiliation{Zhejiang Institute of Modern Physics, Zhejiang University, Hangzhou 310027, China}
\affiliation{CAS Center for Excellence in Topological Quantum Computation, University of Chinese Academy of Sciences, Beijing 100190, China}

\begin{abstract}
We investigate the effect of surface disorder on the chiral surface states of a three-dimensional quantum Hall system.
Utilizing a transfer-matrix method, we find that the localization length of the surface state along the magnetic field decreases with the surface disorder strength in the weak disorder regime, but increases anomalously in the strong disorder regime.
In the strong disorder regime, the surface states mainly locate at the first inward layer to avoid the strong disorder in the outmost layer.
The anomalous increase of the localization length can be explained by an effective model, which maps the strong disorder on the surface layer to the weak disorder on the first inward layer.
Our work demonstrates that surface disorder can be an effective way to control the transport behavior of the surface states along the magnetic field.
We also investigate the effect of surface disorder on the full distribution of conductances $P(g)$ of the surface states in the quasi-one-dimensional (1D) regime for various surface disorder strengths.
In particular, we find that $P(g)$ is Gaussian in the quasi-1D metal regime and log-normal in the quasi-1D insulator regime.
In the crossover regime, $P(g)$ exhibits highly nontrivial forms, whose shapes coincide with the results obtained from the Dorokhov-Mello-Pereyra-Kumar equation of a weakly disordered quasi-1D wire in the absence of time-reversal symmetry.
Our results suggest that $P(g)$ is fully determined by the average conductance, independent of the details of the system, in agreement with the single-parameter scaling hypothesis.

\end{abstract}

\date{\today}

\maketitle

%%%%%%%%%%%%%%%%%%%%%%%%%%%%%%%%%%%%%%%%%%%%%%%%%%%%%%%%%%%%%%%%%%%%%%%%%%%%%%%%%%%%%%%%%%%%%%%%%%%%
\section{Introduction}
\label{sec:intro}
%%%%%%%%%%%%%%%%%%%%%%%%%%%%%%%%%%%%%%%%%%%%%%%%%%%%%%%%%%%%%%%%%%%%%%%%%%%%%%%%%%%%%%%%%%%%%%%%%%%%
The quantum Hall effect (QHE) in two-dimensional (2D) electron systems originates from discrete Landau levels forming under a strong perpendicular magnetic field~\cite{Das97,Prange12}.
In three-dimensional (3D) systems, the band dispersion along the magnetic field ($z$ axis) usually closes the quantum Hall gap.
However, if the interlayer coupling is small compared with the Landau level spacing, we expect that the QHE still exists~\cite{Stormer86,Chalker95}.
This idea was realized in an engineered multilayer quantum well system~\cite{Stormer86,Druist98} and, very recently, in an anisotropic layered material, $\text{BaMnSb}_\text{2}$~\cite{Liu19,Sakai20}.
Even if the interlayer coupling is large enough to close the quantum Hall gap, a gap may further be induced by a spontaneous charge density wave in the $z$ direction under a strong magnetic field~\cite{Halperin87}.
The 3D QHEs recently observed in $\text{ZrTe}_\text{5}$~\cite{Tang19,Galeski20-1,Qin20} and $\text{HfTe}_\text{5}$~\cite{Wang20,Galeski20-2} are suggested to be of this type.
Signatures of 3D QHE have also been found in  Bechgaard salts~\cite{Cooper89,Hannahs89}, $\eta$-$\text{Mo}_{\text{4}}\text{O}_{\text{11}}$~\cite{Hill98}, graphite~\cite{Kopelevich03,Bernevig07}, $n$-doped $\text{Bi}_\text{2}\text{Se}_\text{3}$~\cite{Cao12}, and $\text{EuMnBi}_\text{2}$~\cite{Masuda16}.
 These materials offer us great opportunities to study the QHE beyond two dimensions~\cite{Avron83,Montambaux90,Kohmoto92,Koshino01,Koshino02,Koshino03,Koshino02-2,Koshino04}.

The distinct feature of a 2D quantum Hall system is its chiral edge states, which are topologically protected by the bulk gap and robust against disorder.
In the 3D case, the chiral edge state of each layer is coupled to neighboring edge states, forming a 2D chiral surface state~\cite{Balents96}.
The transport properties of the chiral surface states turn out to be highly anisotropic in the presence of disorder.
Due to the chiral nature of the surface states, the in-plane transport is ballistic.
In the vertical direction, interestingly, there exist three distinct regimes in a mesoscopic sample, namely, 2D chiral metal, quasi-one-dimensional (1D) metal, and quasi-1D insulator~\cite{Mathur97,Balents97,Gruzberg97-1,Gruzberg97-2,Cho97,Plerou98}.

The existence of the 2D chiral surface states was confirmed in Refs.~\cite{Druist98,Druist99,Liu19}.
So far, however, the three transport regimes of the surface states have not been investigated in experiments.
Disorder is inevitable in real experiments.
With the improvement of sample quality, disorder in the bulk tends to be weak, and surface disorder tends to dominate transport.
The latter can be caused by the defects on the surface and adsorption of residual atoms in the vacuum rest gas.
In addition, surface disorder can be easily controlled by adatom deposition, ion sputtering, and air exposure, hence allows a systematic study in experiments.
Therefore, it is important to investigate the effect of surface disorder on the 2D chiral surface states.
Theoretically, the disordered surface states have been mainly investigated using a 2D directed network model~\cite{Chalker95,Kim96,Gruzberg97-1,Gruzberg97-2,Cho97,Plerou98} and a 2D continuum model~\cite{Balents96,Mathur97,Balents97,Betouras00,Tomlinson05-1,Tomlinson05-2}.
However, we note that the scattering strength of an adatom on a sample surface can easily be of the order of magnitude of 1 eV~\cite{Ternes04,Alpichshev12}, which can be much larger than the 3D quantum Hall gap~\cite{Liu19}.
In such a condition, the disorder-induced coupling between the surface and bulk states has to be taken into consideration.
Moreover, as we can see below, the strong disorder on the surface layer tends to push the surface states inward into the bulk.
A 2D model that describes the surface alone cannot capture the physics above, and one needs to treat a full 3D Hamiltonian here. 

In this paper, we study the effect of surface disorder on the 2D chiral surface states using a 3D tight-binding lattice model~\cite{Wang99,Zheng20}.
Surface disorder has also been considered in the context of topological insulators~\cite{ye11,Schubert12,Ringel12,Wu13,Wu14,Sacksteder15,Kim15,Queiroz16}.
Different from topological insulators, our 3D quantum Hall system can be viewed as 2D Chern insulators stacked in the $z$ direction.
While the in-plane surface transport is expected to be topologically protected, there is no such topological protection in the vertical direction.
For 2D topological insulators, it has been shown that the surface conductance is always quantized and remains unchanged by surface disorder at any disorder strength~\cite{Wu13,Kim15}.
Therefore, we focus on vertical transport in this work.
Utilizing a transfer-matrix method, we first determine the localization length of the surface states in the $z$ direction under various surface disorder strengths.
As the disorder strength increases, the localization length decreases in the weak disorder regime but increases anomalously in the strong disorder regime.
The main weight of the surface state gradually moves from the outmost layer to the first inward layer, and finally forms a weakly disordered surface state beneath the disordered surface layer in the large disorder limit~\cite{Schubert12}.
The anomalous increase of the localization length in the strong disorder regime can be explained by an effective model~\cite{Ringel12}, which maps the strong disorder on the surface to the weak disorder on the first inward layer.
Our results demonstrate that surface disorder can be an effective way to control the transport behavior of the surface states in the $z$ direction.
For the localized surface state in the intermediate disorder regime, the conduction can be further enhanced by doping disorder on its surface, forming a more extended state beneath the outmost disordered layer. 

We also investigate the effect of surface disorder on the conductance distributions $P(g)$ of the chiral surface states in the quasi-1D regime.
The conductance distributions of the chiral surface states have been mainly investigated using the 2D directed network model in the literature~\cite{Gruzberg97-1,Gruzberg97-2,Plerou98}.
Analytically, Gruzberg, Read, and Sachdev proved that in the quasi-1D regime, the conductance properties of the 2D directed network model are the same as those of a weakly disordered quasi-1D wire~\cite{Gruzberg97-1}.
The conductance distributions of the latter can be calculated from the Dorokhov-Mello-Pereyra-Kumar (DMPK) equation of the Fokker-Planck approach~\cite{Dorokhov82,*Mello88,Beenakker94,Beenakker97,Mirlin00,Muttalib99,*Gopar02,*Muttalib03,Froufe02}.
Direct numerical investigations of the $P(g)$ of the chiral surface states are rather limited in the literature.
To the best of our knowledge, the only work was done in Ref.~\cite{Plerou98}, which studied $P(g)$ using the 2D directed network model in the quasi-1D regime.
However, the $P(g)$ in the crossover regime between the quasi-1D metal and insulator regimes was not closely examined in Ref.~\cite{Plerou98}.
We emphasize that both the network model and the DMPK equation are, however, only valid in weakly disordered systems.
It is unclear whether the conductance distributions of the chiral surface states in the quasi-1D regime can still be described by the DMPK equation in the presence of strong surface disorder, which is the typical case in realistic samples.

Using the 3D tight-binding model, we investigate $P(g)$ of the chiral surface states under various surface disorder strengths and provide a detailed study of the $P(g)$ in the crossover regime.
It is found that $P(g)$ is Gaussian in the quasi-1D metal regime and log-normal in the quasi-1D insulator regime as expected.
In the crossover regime, $P(g)$ is found to exhibit highly nontrivial forms, whose shapes coincide with the results obtained from the DMPK equation of a weakly disordered quasi-1D wire in the absence of time-reversal symmetry (unitary universality class)~\cite{Gruzberg97-1,Muttalib99,*Gopar02,*Muttalib03,Froufe02}.
Our results suggest that $P(g)$ is the only function of the average conductance, independent of the surface disorder strength and the size of the system, in agreement with the single-parameter scaling hypothesis.
For most of our work, the disorder is only introduced at the surface of the sample, and the bulk is left clean.
Finally, we examine the effect of weak disorder in the bulk, which is often the case in a realistic sample.

The rest of the paper is organized as follows. In Sec.~\ref{sec:model} we describe the tight-binding Hamiltonian for the 3D quantum Hall system and the numerical method we use. In Sec.~\ref{sec:Results} we present our numerical results. The paper is summarized in Sec.~\ref{sec:conclusions}.

%%%%%%%%%%%%%%%%%%%%%%%%%%%%%%%%%%%%%%%%%%%%%%%%%%%%%%%%%%%%%%%%%%%%%%%%%%%%%%%%%%%%%%%%%%%%%%%%%%%%
\section{Model and Method}
\label{sec:model}
%%%%%%%%%%%%%%%%%%%%%%%%%%%%%%%%%%%%%%%%%%%%%%%%%%%%%%%%%%%%%%%%%%%%%%%%%%%%%%%%%%%%%%%%%%%%%%%%%%%%

%%%%%%%%%%%%%%%%%%%%%%%%%%%%%%%%%%%%%%%%%%%%%%%%%%%%%%%%%%%%%%%%%%%%%%%%%%%%%%%%%%%%%%%%%%%%%%%%%%%%
%\subsection{3D lattice model}
%\label{subsec:lattice_model}
We consider an electron on an $L_x \times L_y \times L_z$ cubic lattice in the presence of a magnetic field $B \hat{z}$ with tight-binding Hamiltonian
\begin{equation}
\mathcal{H}=-\sum_{\langle i,j\rangle}\left (t_{ij} e^{i\theta_{ij}}c_i^\dagger c_j+\mathrm{H.c.}\right),
\label{eq:lattice_model}
\end{equation}
where we have anisotropic nearest-neighboring hopping
\begin{equation*}
t_{ij} = \left \{
\begin{array}{ll}
1 & {\rm \ }i{\rm \ and\ }j{\rm \ are\ horizontal\ nearest\ neighbors,} \\
t_z & {\rm \ }i{\rm \ and\ }j{\rm \ are\ vertical\ nearest\ neighbors,} \\
0 & {\rm \ }i{\rm \ and\ }j{\rm \ are\ not\ nearest\ neighbors.}
\end{array}
\right .
\end{equation*} 
We choose Landau gauge $\vec{A}=(0,Bx,0)$ and define $\theta_{ij}=\frac{e}{\hbar}\int_i^j \vec{A}\cdot \mathrm{d} \vec{l}$.
The magnetic flux $\phi$ per unit cell in a horizontal plane is
\begin{equation}
\frac{\phi}{\phi_0}=\frac{Ba^2}{hc/e}=\frac{1}{2\pi}\sum_{\Box} \theta_{ij},
\end{equation}
where $\phi_0 = hc/e$ is the flux quantum.

In the 2D limit with $L_z = 1$, this model has a butterfly-like self-similar energy spectrum, as the flux $\phi$ per unit cell varies~\cite{Hofstadter76}.
When the flux $\phi$ per unit cell is chosen as $\phi_0 / N$ for integer $N$, there are exactly $N$ subbands in the spectrum.
Here we consider the case where $t_z$ is much smaller than the horizontal hopping $t$ = 1, so that the subband gaps are not closed by the dispersion in the $z$ axis. 
The 2D chiral surface states lie in the gap regions between the subbands and can be revealed by imposing open boundary conditions in the $x$ and $y$ directions.

To study the effect of surface disorder on the surface states, we consider the random on-site potential given by
\begin{equation}
\mathcal{H}_\mathrm{imp}=\sum_i \epsilon_i c_i^\dagger c_i,
\label{eq:lattice_model}
\end{equation}
where $\epsilon_i$ are independent variables with identical uniform distribution on $[-W/2, W/2]$, and $W$ is the surface disorder strength. 
The effect of bulk disorder, where disorder is added on every site of the system, 
on the bulk and surface states of this model has been studied in Refs.~\cite{Wang99,Zheng20}.
Here we consider surface disorder and introduce disorder only for the sites at the outmost sidewalls of the sample.

\begin{figure}
\centering
\includegraphics[width=\columnwidth]{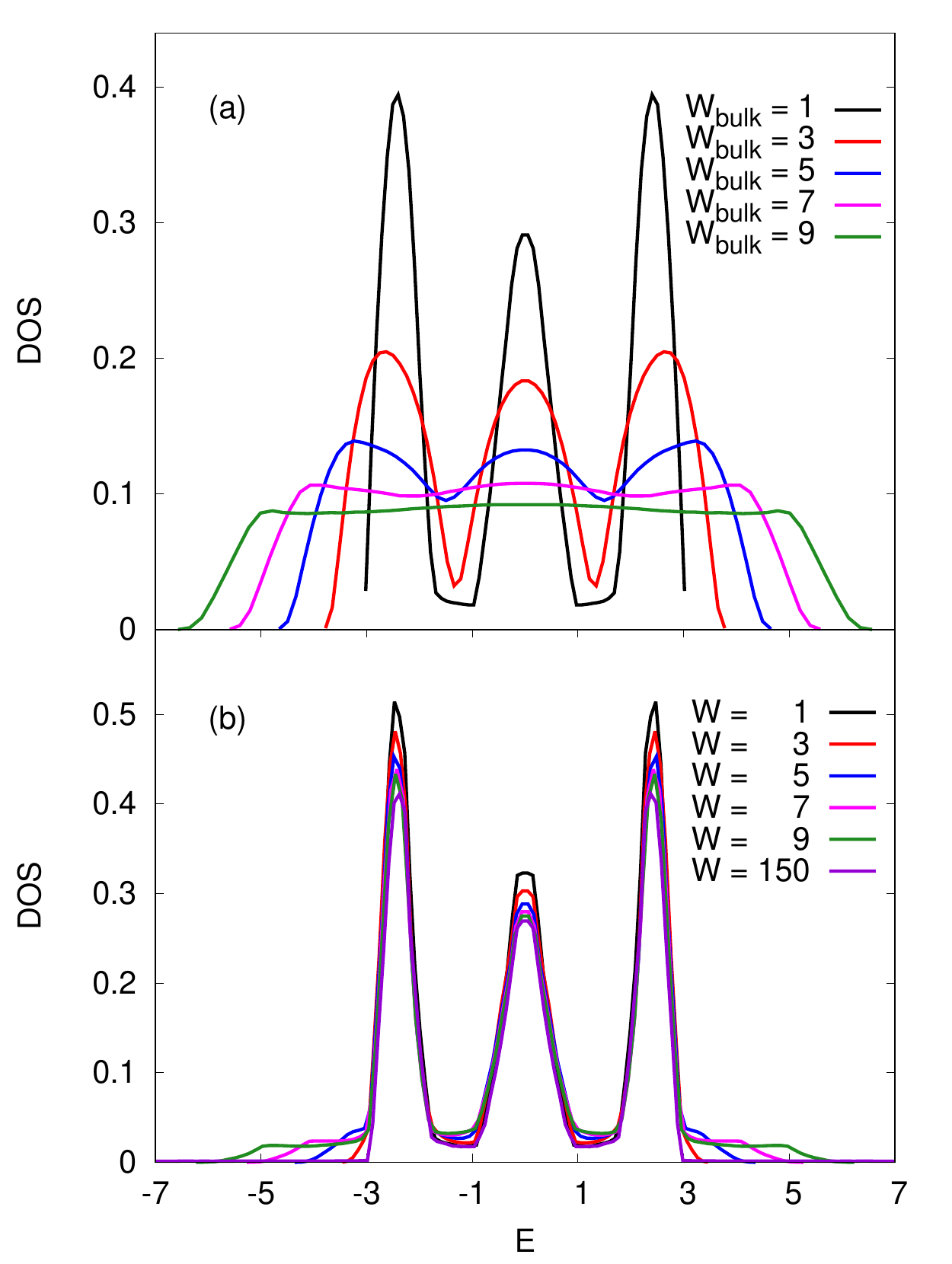}
\caption{Density of states of a $21 \times 21 \times 21$ cubic lattice with $\phi = \phi_0/3$, $t_z = 0.1$ for (a) bulk-disordered systems with disorder strength $W_\mathrm{bulk}$ and (b) surface-disordered systems with disorder strength $W$.
Here we apply OBCs in the $x$ and $y$ directions and a PBC in the $z$ direction.}
\label{fig:dos}
\end{figure}

The evolution of the density of states (DOS) as a function of the disorder strength for both bulk-disordered and surface-disordered systems is illustrated in Fig.~\ref{fig:dos}. 
Here we choose $\phi = \phi_0/3$, $t_z = 0.1$. 
The system size is $21 \times 21 \times 21$, and we apply open boundary conditions (OBCs) in the $x$ and $y$ directions and a periodic boundary condition (PBC) in the $z$ direction.
In such a case, the 2D chiral surface states can be identified by the finite DOS inside the bulk band gaps.
For bulk-disordered systems, the Landau bands broaden as the bulk disorder strength $W_\mathrm{bulk}$ increases.
The gaps between the subbands close at about $W_\mathrm{bulk} = 3$.
For $W_\mathrm{bulk} > 3$, the three Landau bands merge into a single large band, which keeps broadening with increasing $W_\mathrm{bulk}$.
The DOS of surface-disordered systems displays quite different behaviors, as shown in Fig.~\ref{fig:dos}(b).
Similar to the bulk-disordered case, the band tails of the DOS expand as the surface disorder strength $W$ increases.
The central region, however, is relatively stable against disorder.
The finite DOS of the surface states inside the band gaps persists even when $W = 150$, indicating that the surface states are topologically protected by the bulk gap and cannot be destroyed by surface disorder.

To calculate the two-terminal vertical conductance, we attach two semi-infinite clean leads at the top and bottom ends of the sample.
The conductance is calculated from the Landauer-B\"uttiker formula~\cite{Landauer57,*Economou81,*Fisher81}
\begin{equation}
 G = \frac{2e^2}{h} \operatorname{Tr}(\boldsymbol{t} \boldsymbol{t}^\dagger).
\label{eq:Landauer-Buttiker}
\end{equation}
In it, $\boldsymbol{t}$ is the transmission matrix, which we calculate using the transfer-matrix method~\cite{Pendry92,Markos06}.
For simplicity, we use the dimensionless conductance $g$ defined as $g = G/(2e^2/h)$ in the rest of the paper.

%%%%%%%%%%%%%%%%%%%%%%%%%%%%%%%%%%%%%%%%%%%%%%%%%%%%%%%%%%%%%%%%%%%%%%%%%%%%%%%%%%%%%%%%%%%%%%%%%%%%
\section{Results}
\label{sec:Results}
%%%%%%%%%%%%%%%%%%%%%%%%%%%%%%%%%%%%%%%%%%%%%%%%%%%%%%%%%%%%%%%%%%%%%%%%%%%%%%%%%%%%%%%%%%%%%%%%%%%%

%%%%%%%%%%%%%%%%%%%%%%%%%%%%%%%%%%%%%%%%%%%%%%%%%%%%%%%%%%%%%%%%%%%%%%%%%%%%%%%%%%%%%%%%%%%%%%%%%%%%
\subsection{Localization length}
\label{subsec:Localization_length}
%%%%%%%%%%%%%%%%%%%%%%%%%%%%%%%%%%%%%%%%%%%%%%%%%%%%%%%%%%%%%%%%%%%%%%%%%%%%%%%%%%%%%%%%%%%%%%%%%%%%

Due to the chiral nature of the edge states, the transport of the surface states is ballistic in the $x$-$y$ plane.
Furthermore, the unidirectional transport in the  $x$-$y$ plane suppresses the localization effect in the $z$ direction.
In order to make quantum interference happen, an electron has to circumnavigate the sample and return to its starting point.
This is impossible in an infinite sample.
Thus, for an infinite sample, vertical transport is always diffusive, independent of the disorder strength~\cite{Balents96}.

In a mesoscopic sample, an electron can circle the sample and interfere with itself.
For a very long length $L_z$, the system is of quasi-1D nature.
The interference can happen many times so that the surface state is localized in the $z$ direction.
This is the so-called quasi-1D insulator regime of the chiral surface states.
For $l \ll L_z \ll \xi$, where $l$ is the mean free path and $\xi$ is the localization length, the system is in the diffusive regime.
Here another characteristic length scale emerges and separates the diffusive regime into two regimes~\cite{Gruzberg97-2}.
During one round-trip of the sample, an electron diffuses a distance $L_0$ in the vertical direction.
If $L_0 \ll L_z \ll \xi$, that means the electron circles around the sample many times before diffusing out, and the system is in the quasi-1D metal regime.
If $l \ll L_z \ll L_0$, the electron diffuses out of the sample without a complete round-trip, and the system is in the 2D chiral metal regime.
In terms of the average conductance, both regimes share the same Ohmic behavior.
However, the conductance fluctuations can be much larger in the 2D chiral metal regime, since the system can be effectively broken up into independent parallel strips, whose width is the distance an electron propagates in the chiral direction during the trip~\cite{Mathur97,Gruzberg97-2}.
We note that to avoid entering into the ballistic regime, the system size needed for the 2D chiral metal regime is rather large for a 3D tight-binding Hamiltonian~\cite{Cho97,Plerou98,Zheng20}.
Therefore, we focus on the quasi-1D metal and insulator regimes in this paper.

\begin{figure}
\centering
\includegraphics[width=\columnwidth]{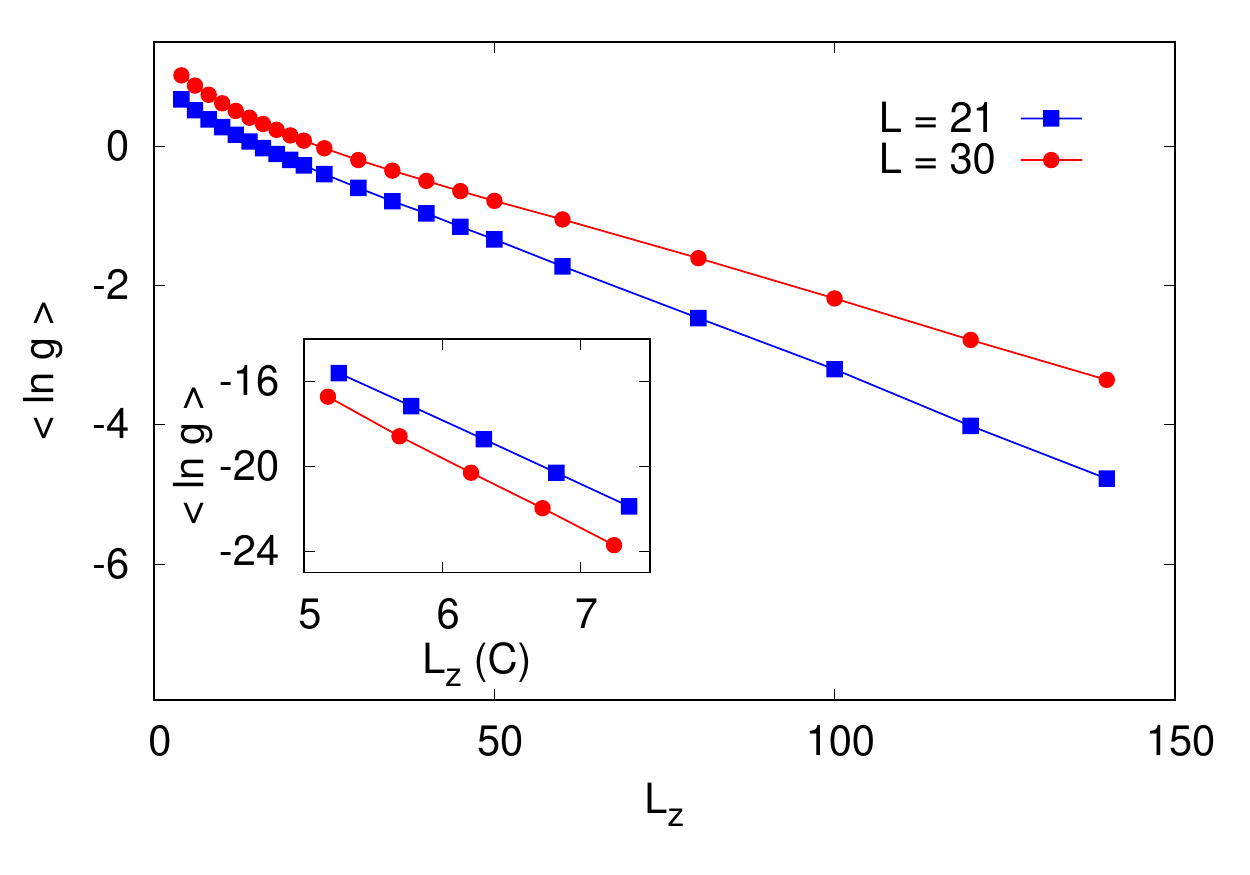}
\caption{Averaged logarithm of conductance $\langle \ln g \rangle$ as a function of the length $L_z$ in a quasi-1D system $L \times L \times L_z$ at $E = -1.35$ for widths $L = 21$ and 30.
Here $\phi = \phi_0/3$, $t_z = 0.1$, and the surface disorder strength $W = 1$.
The average is taken over $10^4$ disorder realizations.
The inset shows $\langle \ln g \rangle$ as a function of $L_z$ in units of $C$ in the insulating regime, where $C$ is the circumference of the sample.
We determine the localization length from the inset using $\langle \ln g \rangle = -2L_z/\xi$. 
The fitting yields $\xi=53.5 \pm 0.1$ for $L =21$ and $\xi=69.1 \pm 0.7$ for $L =30$.}
\label{fig:lng}
\end{figure}

The characteristic length scale that separates the quasi-1D metal and insulator regimes is the localization length $\xi$.
First, we study how the surface disorder strength affects the localization length of the surface states in the $z$ direction.
The localization length can be determined from the scaling behavior of the average conductance.
For relatively short samples, the average conductance follows a typical Ohmic behavior.
For relatively long samples, the average conductance decays exponentially with the length $L_z$ in the form
\begin{equation}
\langle \ln g \rangle \sim -\frac{2L_z}{\xi}.
\label{eq:localization_length}
\end{equation}
The crossover from the quasi-1D metal to insulator regime occurs at $\langle g \rangle \sim 1$, where $L_z$ is of the order of $\xi$.
Figure~\ref{fig:lng} shows $\langle \ln g \rangle$ as a function of $L_z$ in a quasi-1D system $L \times L \times L_z$ for two different transverse system sizes $L = 21$ and 30.
Here we choose $\phi = \phi_0/3$, $t_z = 0.1$, and the surface disorder strength $W = 1$.
The energy is at $E = -1.35$, which is near the center of the subband gap (see Fig.~\ref{fig:dos}).
For a quasi-1D system, the localization length is expected to be proportional to the number of conducting channels $N$~\cite{Beenakker97}.
Since the number of conducting channels of the surface states is proportional to the circumference $C$ of the sample, we expect the localization length to be approximately proportional to $C$.
This is verified in the inset of Fig.~\ref{fig:lng}, where $\langle \ln g \rangle$ is plotted as a function of $L_z$ in units of $C$ for very long samples.
The two straight lines are almost parallel to each other.
We obtain the localization lengths from the inset by using Eq.~(\ref{eq:localization_length}), and the fitting yields $\xi=53.5 \pm 0.1$ for $L =21$ and $\xi=69.1 \pm 0.7$ for $L =30$.

\begin{figure}
\centering
\includegraphics[width=\columnwidth]{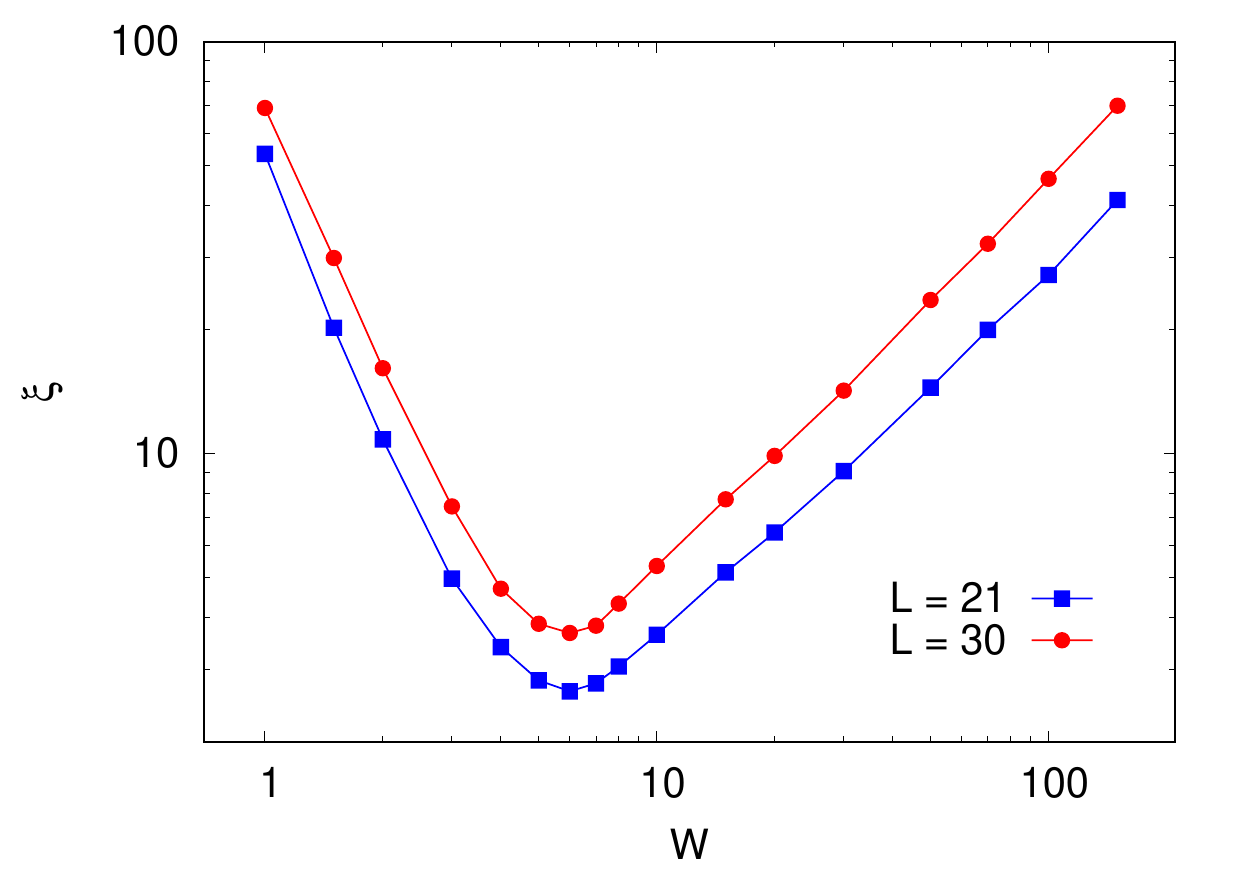}
\caption{Log-log plot of the localization length $\xi$ as a function of the surface disorder strength $W$ in a quasi-1D system $L \times L \times L_z$ at $E = -1.35$ for $L = 21$ and 30.
Here $\phi = \phi_0/3$, $t_z = 0.1$.
Both curves reach their minimum at $W_c = 6$, and the distance between them becomes larger in the strong disorder regime.}
\label{fig:kesi_vs_W}
\end{figure}

By repeating the above procedure, we calculate the localization length as a function of the surface disorder strength in Fig.~\ref{fig:kesi_vs_W}.
For both widths $L$, the localization length decreases as the disorder strength increases at weak disorder.
However, after a critical disorder strength $W_c$, which is of the order of the bandwidth, the localization length increases anomalously with the disorder strength.
In other words, the conduction of the surface state is enhanced by surface disorder in this regime.
Moreover, we note that the distance between the two curves becomes slightly larger in the strong disorder regime.

\begin{figure*}
\centering
\includegraphics[width=2.1\columnwidth]{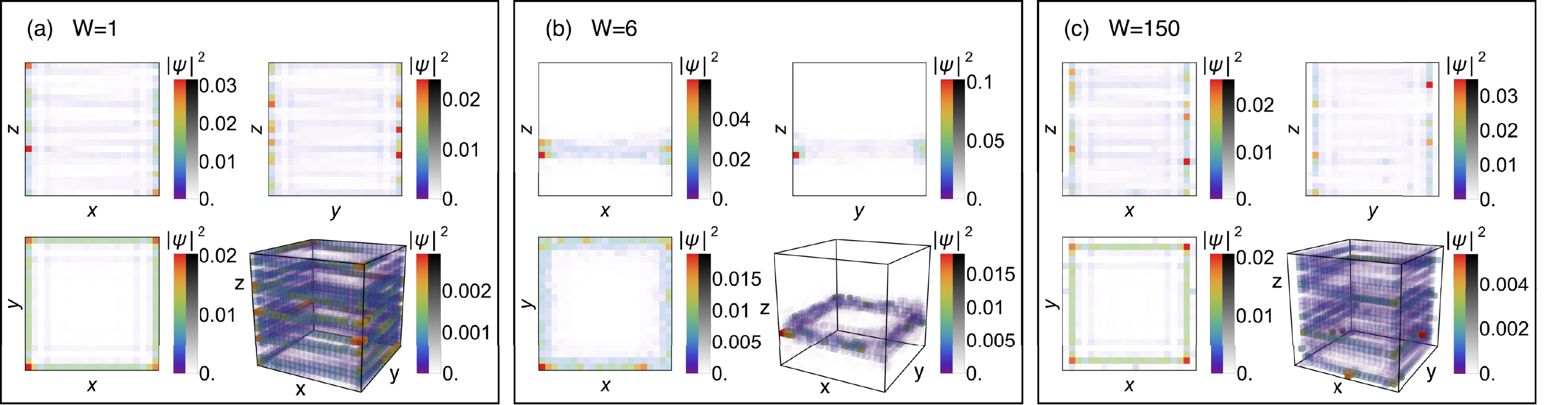}
\caption{Typical surface states in a $21 \times 21 \times 21$ cubic lattice at $E = -1.35$ for surface disorder strengths (a) $W = 1$, (b) $W = 6$, and (c) $W = 150$. 
Here $\phi = \phi_0/3$, $t_z = 0.1$.
The results are obtained by exact diagonalization under OBCs in the $x$ and $y$ directions and a PBC in the $z$ direction for three particular disorder realizations.
We show both the plots of the 3D probability density $\left| \psi \right|^2$ and its projections onto the $x$-$y$, $x$-$z$, and $y$-$z$ planes.
Each lattice point is represented by a small cube (square), whose color and opacity depend on the value of $\left| \psi_i \right|^2$.
The color and opacity bar is given on the right of each plot.
For weak disorder $W = 1$, the surface state mainly locates at the outermost layer and is extended in the $z$ direction.
At intermediate disorder $W = 6$, the surface state moves inward significantly, and it becomes inhomogeneous in the $x$-$y$ plane and localized in the $z$ direction.
For very strong disorder $W = 150$, the surface state mainly locates at the first inward layer to avoid the strong disorder in the outmost layer, and it becomes extended again in the $z$ direction.}
\label{fig:wavefunction}
\end{figure*}

\begin{figure}
\centering
\includegraphics[width=\columnwidth]{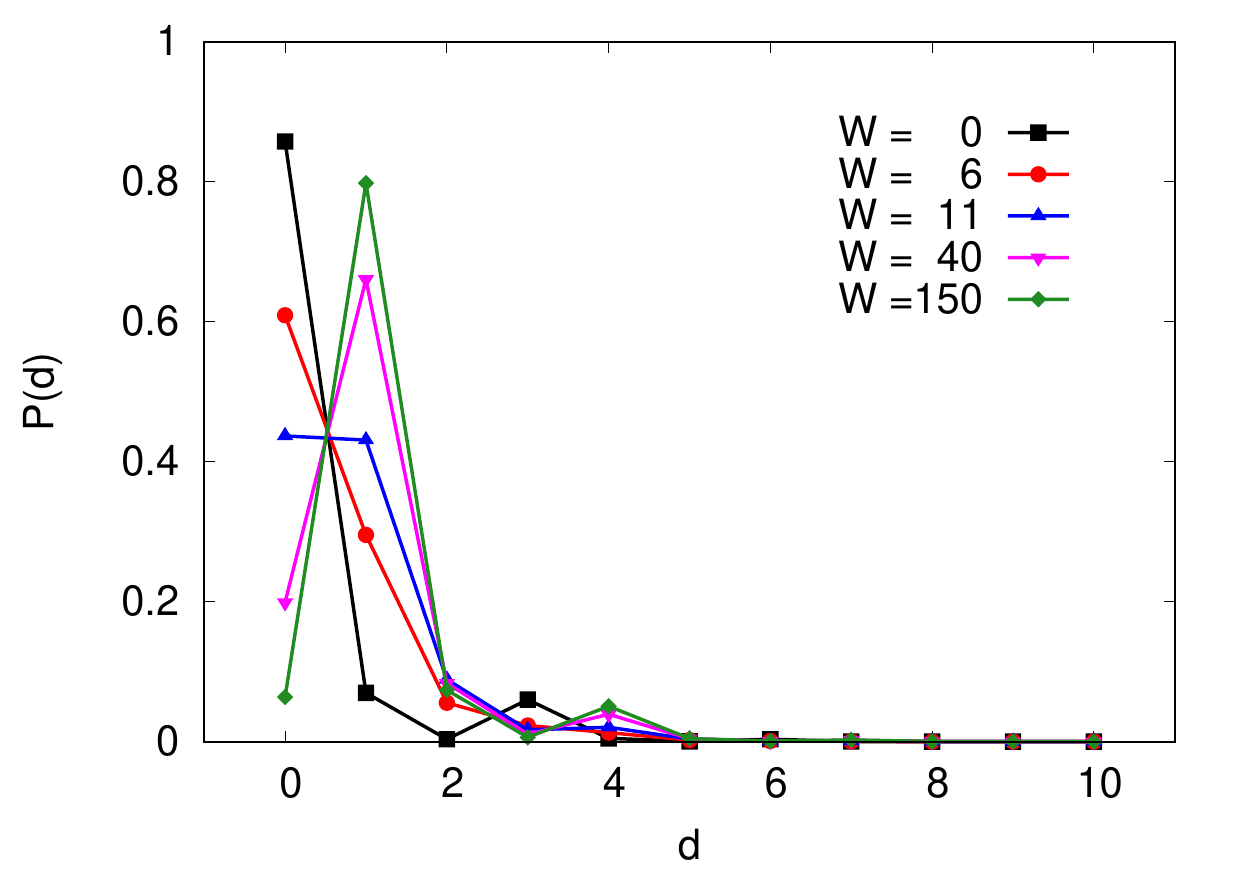}
\caption{The probability $P(d)=\int d^3 \vec{x}|\psi(\vec{x})|^2 \delta[d-d(\vec{x})]$ as a function of distance $d$ from the surface in a $21 \times 21 \times 21$ cubic lattice  at $E = -1.35$ for surface disorder strengths $W = 0$, 6, 11, 40, and 150.
Here $\phi = \phi_0/3$, $t_z = 0.1$. 
The average is taken over $10^3$ disorder realizations.
As the surface disorder strength increases, the main weight of the surface state gradually moves from the outmost layer to the first inward layer.}
\label{fig:Pd}
\end{figure}

To understand the anomalous increase of the localization length in the strong disorder regime, in Fig.~\ref{fig:wavefunction} we plot the typical surface states in a $21 \times 21 \times 21$ cubic lattice at $E = -1.35$ for surface disorder strengths $W = 1$, 6, and 150.
For weak disorder $W = 1$, the surface state mainly locates at the outermost layer and is extended in the $z$ direction.
At intermediate disorder $W = 6$, the surface state moves inward significantly, and it becomes inhomogeneous in the $x$-$y$ plane and localized in the $z$ direction.
For very strong disorder $W = 150$, the surface state mainly locates at the first inward layer and becomes extended again in the $z$ direction.
The surface states in the $x$-$y$ plane are topologically protected by the bulk gap of the system.
Since surface disorder does not alter the bulk gap, it never destroys the surface states in the $x$-$y$ plane.
This is also verified in the DOS in Fig.~\ref{fig:dos}(b).
For very strong disorder, the surface layer becomes an Anderson insulator~\cite{Schubert12}.
The redistributed surface state on the first inward layer can be considered as an interface state between an Anderson insulator and a 3D quantum Hall system~\cite{Schubert12,Ringel12}.
More quantitatively, in Fig.~\ref{fig:Pd}, we plot the probability $P(d)=\int d^3 \vec{x}|\psi(\vec{x})|^2 \delta[d-d(\vec{x})]$ as a function of distance $d$ from the surface in a $21 \times 21 \times 21$ cubic lattice  at $E = -1.35$ for $W = 0$, 6, 11, 40, and 150.
As the surface disorder strength increases, the main weight of the surface state gradually moves from the outmost layer to the first inward layer.

The larger distance between the curves of localization lengths in the strong disorder regime in Fig.~\ref{fig:kesi_vs_W} can be understood from the movement of the surface states in the $x$-$y$ plane.
Due to the double-log plot in Fig.~\ref{fig:kesi_vs_W}, the distance between the curves measures the ratio of the localization lengths.
As mentioned above, the localization length is expected to be proportional to the circumference $C$ of the surface states.
In the weak disorder regime, the surface states mainly locate at the outermost layer, and the ratio of $C$ at $L = 30$ to $C$ at $L = 21$ can be calculated as $(30 \times 4-4)/(21 \times 4-4)=1.45$.
On the other hand, in the strong disorder regime, the surface states mainly locate at the first inward layer, and the $C$ ratio increases to $(28 \times 4-4)/(19 \times 4-4)=1.5$.
This explains why the ratio of the localization lengths becomes larger in the strong disorder regime.

Since the surface states mainly locate at the first inward layer in the strong disorder regime, one can derive an effective model that describes the effect of surface disorder on the first inward layer~\cite{Ringel12}.
To proceed, we divide the system into two parts: the clean bulk and the disordered surface layer.
The Schr\"odinger equation of the whole system can be written as
\begin{equation}
\left(\begin{array}{cc}H_0 & V \\ V^\dagger & H_\mathrm{dis}\end{array}\right)\left(\begin{array}{c}\psi_0 \\ \psi_\mathrm{dis}\end{array}\right)=E\left(\begin{array}{c}\psi_0 \\ \psi_\mathrm{dis}\end{array}\right),
\label{eq:Schrodinger_equation}
\end{equation}
where $H_0$ is the Hamiltonian for the clean bulk, $H_\mathrm{dis}$ is the Hamiltonian for the disordered surface layer, $V$ and $V^\dagger$ are the couplings between them, and $\psi_0$ and $\psi_\mathrm{dis}$ are the corresponding wave functions.
In the strong disorder regime, $W \gg t$, $\psi_\mathrm{dis}$ may be considered as a high-energy sector and can be integrated out.
Eliminating $\psi_\mathrm{dis}$ in Eq.~(\ref{eq:Schrodinger_equation}), we obtain an effective Hamiltonian for the clean bulk
\begin{equation}
\left(H_{0}-\frac{V V^\dagger}{H_\mathrm{dis}-E}\right) \psi_0=E \psi_0,
\label{eq:effetive_model}
\end{equation}
which means that the disorder potential on the first inward layer is renormalized into $V(H_\mathrm{dis}-E)^{-1}V^\dagger$.
Physically, this term describes the virtual hopping from the clean bulk to the states on the disordered surface layer with typically very different energies, and finally back to the bulk.
Since the matrix elements of $V$ are of the order of $t$, the effective disorder on the first inward layer is of the order of $t^2/W$, which is much weaker than the disorder strength $W$ on the outmost layer.
As $W$ increases, the effective disorder on the first inward layer decreases.
This explains the anomalous increase of the localization length in the strong disorder regime.

The above argument suggests that the minimum of the localization length $W_c$  is situated at disorder strength of the order of the in-plane hopping strength $t$ and is independent of the interlayer hopping strength $t_z$.
This is indeed the case in the numerical simulation in Fig.~\ref{fig:kesi_vs_W}, where $W_c = 6$ for $t = 1$, $t_z = 0.1$.
To further verify this point, we have numerically checked that when the in-plane hopping strength doubles to $t = 2$, $W_c$ also doubles to 12; on the other hand, $W_c$ remains unchanged when the interlayer hopping strength $t_z$ is reduced to 0.01. 

%%%%%%%%%%%%%%%%%%%%%%%%%%%%%%%%%%%%%%%%%%%%%%%%%%%%%%%%%%%%%%%%%%%%%%%%%%
%%%%%%%%%%%%%%%%%%%%%%%%%%
\subsection{Conductance distributions}
\label{subsec:Conductance_distributions}
%%%%%%%%%%%%%%%%%%%%%%%%%%%%%%%%%%%%%%%%%%%%%%%%%%%%%%%%%%%%%%%%%%%%%%%%%%
%%%%%%%%%%%%%%%%%%%%%%%%%%

So far, we have investigated the effect of surface disorder on the localization length of the chiral surface states in the $z$ direction, which can be determined from the scaling of the average conductance.
In the following, we consider the effect of surface disorder on the full distribution of the conductances in the quasi-1D regime.

The conductance distributions of the chiral surface states have been mainly investigated using a 2D directed network model in the literature~\cite{Gruzberg97-1,Gruzberg97-2,Plerou98}.
Analytically, Gruzberg, Read, and Sachdev proved that in the quasi-1D regime, the conductance properties of the 2D directed network model are the same as those of a weakly disordered quasi-1D wire~\cite{Gruzberg97-1}.
The latter has been extensively investigated, and a nearly complete description of the conductance properties is available in the literature~\cite{Beenakker97,Mirlin00}.
The first two moments of the conductance distribution have been calculated using the supersymmetric nonlinear $\sigma$ model~\cite{Zirnbauer92,Mirlin94}.
Furthermore, the full probability distribution of the transmission eigenvalues $P(\lbrace T_n \rbrace)$ can be obtained from the DMPK equation of the Fokker-Planck approach~\cite{Dorokhov82,*Mello88}.
The DMPK equation describes the evolution of $P(\lbrace \lambda_n \rbrace)$ with increasing wire length $L_z$~\cite{Beenakker97}:

\begin{align}
&l \frac {\partial P}{\partial L_z} = \frac{2}{\beta N+2-\beta} \sum_{n=1}^N \frac{\partial}{\partial \lambda_n} \lambda_n (1+\lambda_n) J \frac{\partial}{\partial \lambda_n} \frac{P}{J}, \\
&J=\prod_{i=1}^N \prod_{j=i+1}^N|\lambda_j-\lambda_i|^\beta,
\label{eq:DMPK_equation}
\end{align}
where $\lambda_n$ is related to $T_n$ by $\lambda_n = (1-T_n)/T_n$, and $\beta$ is the symmetry index, $\beta = 1$, 2, or 4 for orthogonal, unitary, or symplectic class, respectively.
For the unitary class, which is the case we study here, the DMPK equation can be exactly solved~\cite{Beenakker94}.
The conductance distribution $P(g)$ can be further calculated from $P(\lbrace T_n \rbrace)$, which was performed in Refs.~\cite{Muttalib99,*Gopar02,*Muttalib03,Froufe02}.
Thus, the equivalence between the two models offers us great insight into the conductance properties of the chiral surface states in the quasi-1D regime.
Numerical investigations of $P(g)$ of the chiral surface states are rather limited in the literature.
To the best of our knowledge, the only work was done in Ref.~\cite{Plerou98}, which studied $P(g)$ using the 2D directed network model in the quasi-1D regime.
However, the conductance distributions in the crossover regime between the quasi-1D metal and insulator regimes were not closely examined in Ref.~\cite{Plerou98}.

It is important to note that both the network model and the DMPK equation are only valid in weakly disordered systems.
The network model describes the percolation of electrons in a strong magnetic field and smooth disorder potential~\cite{Chalker88,Kramer05}.
In the presence of strong surface disorder, it is no longer valid and cannot capture the inward movement of the surface states.
The derivation of the DMPK equation is also based on the assumption that the wire is weakly disordered, so that the scattering in each increasing step $\delta L_z$ can be treated perturbatively ~\cite{Mello88,Brouwer96,Beenakker97,Markos06}.
Thus, it is an open question whether the conductance distributions of the chiral surface states in the quasi-1D regime can still be described by the DMPK equation in the presence of strong surface disorder, which is the typical case in realistic samples.
In the following, we investigate the effect of surface disorder on the conductance distributions of the surface states using the 3D tight-binding model under various disorder strengths. 

\begin{figure}
\centering
\includegraphics[width=\columnwidth]{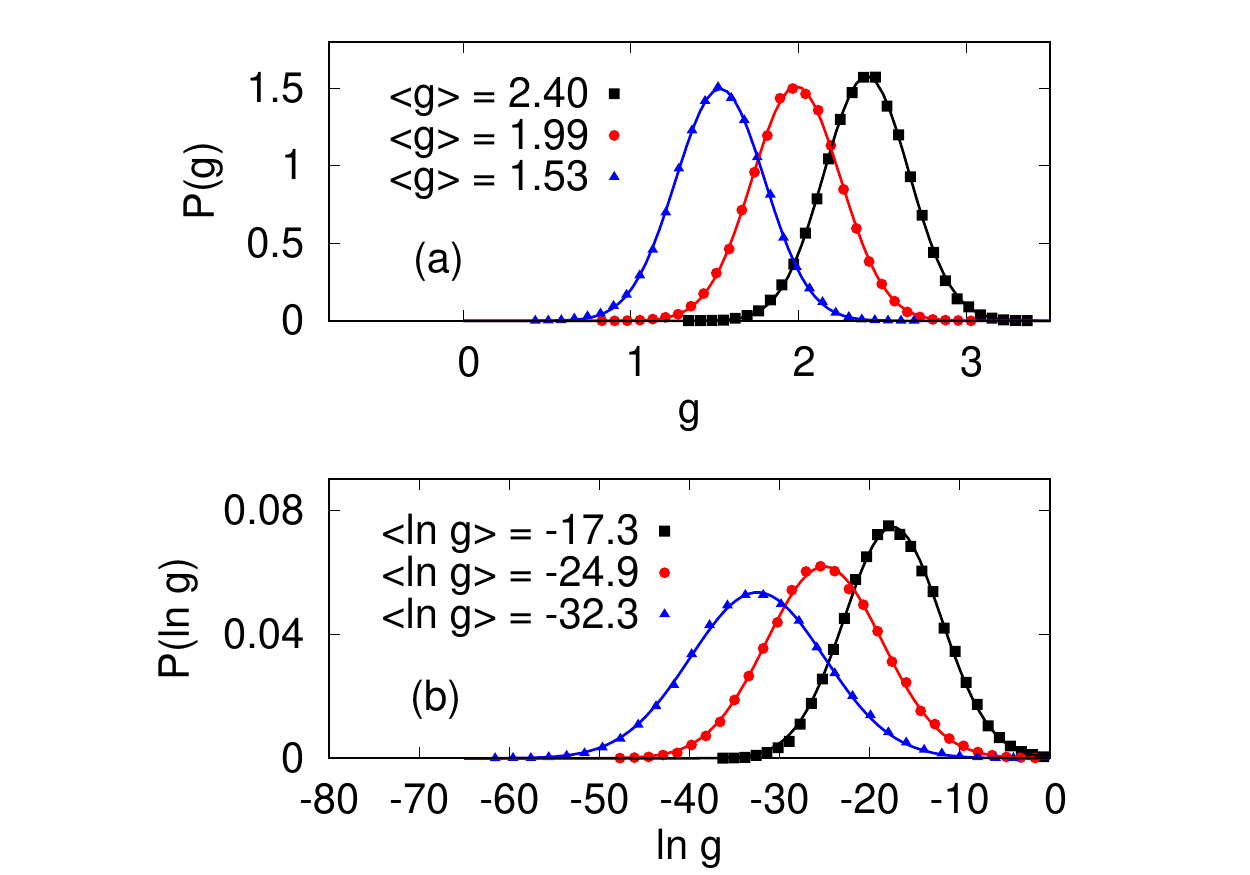}
\caption{Conductance distributions (a) $P(g)$ and (b) $P(\ln g)$ in a quasi-1D system $L \times L \times L_z$ at $E = -1.35$ in the (a) quasi-1D metal regime and (b) quasi-1D insulator regime. 
Here $\phi = \phi_0/3$, $t_z = 0.1$.
To construct each histogram, $5 \times 10^4$ disorder realizations were used.
Solid lines are fit to (a) Gaussian and (b)  log-normal distributions.
For (a), we use $L = 30$ and $W = 1$, and the curves from right to left correspond to $L_z = 6$, 9, and 14;
for (b), $L = 21$ and $W = 6$, and the curves from right to left correspond to $L_z = 25$, 35, and 45.}
\label{fig:Pg_diffusive_localized}
\end{figure}

We first present the results in the quasi-1D metal regime.
Figure~\ref{fig:Pg_diffusive_localized}(a) shows the conductance distributions $P(g)$ in a quasi-1D system $30 \times 30 \times L_z$ at $E = -1.35$ with $\phi = \phi_0/3$, $t_z = 0.1$, and $W = 1$.
We recall that the localization length of this system is $\xi=69.1 \pm 0.7$, which was calculated in Sec.~\ref{subsec:Localization_length}.
For $L_z = 6$, 9, and 14, $L_z \ll \xi$, and the system is deeply in the metallic regime.
As shown in the figure, $P(g)$ is well approximated by a Gaussian in this regime.
We note that the widths of the distributions barely change with $L_z$ at small $L_z/\xi$.
The variance of the conductance is 0.0637, 0.0692, and 0.0715 for $L_z = 6$, 9, and 14, respectively, which is close to the universal value 1/15 in the unitary class~\cite{Gruzberg97-2,Cho97,Plerou98}.

For the quasi-1D insulator regime, in Fig.~\ref{fig:Pg_diffusive_localized}(b) we plot the conductance distributions $P(\ln g)$ in a quasi-1D system $21 \times 21 \times L_z$ at $E = -1.35$ with $\phi = \phi_0/3$, $t_z = 0.1$, and $W = 6$.
The calculated localization length is $\xi=2.65 \pm 0.01$ for this system.
We choose $L_z = 25$, 35, and 45, which fulfills $L_z \gg \xi$, to plot the conductance distributions.
 As shown in the figure, $P(\ln g)$ can be well fitted by log-normal distributions in this regime.

\begin{figure}
\centering
\includegraphics[width=\columnwidth]{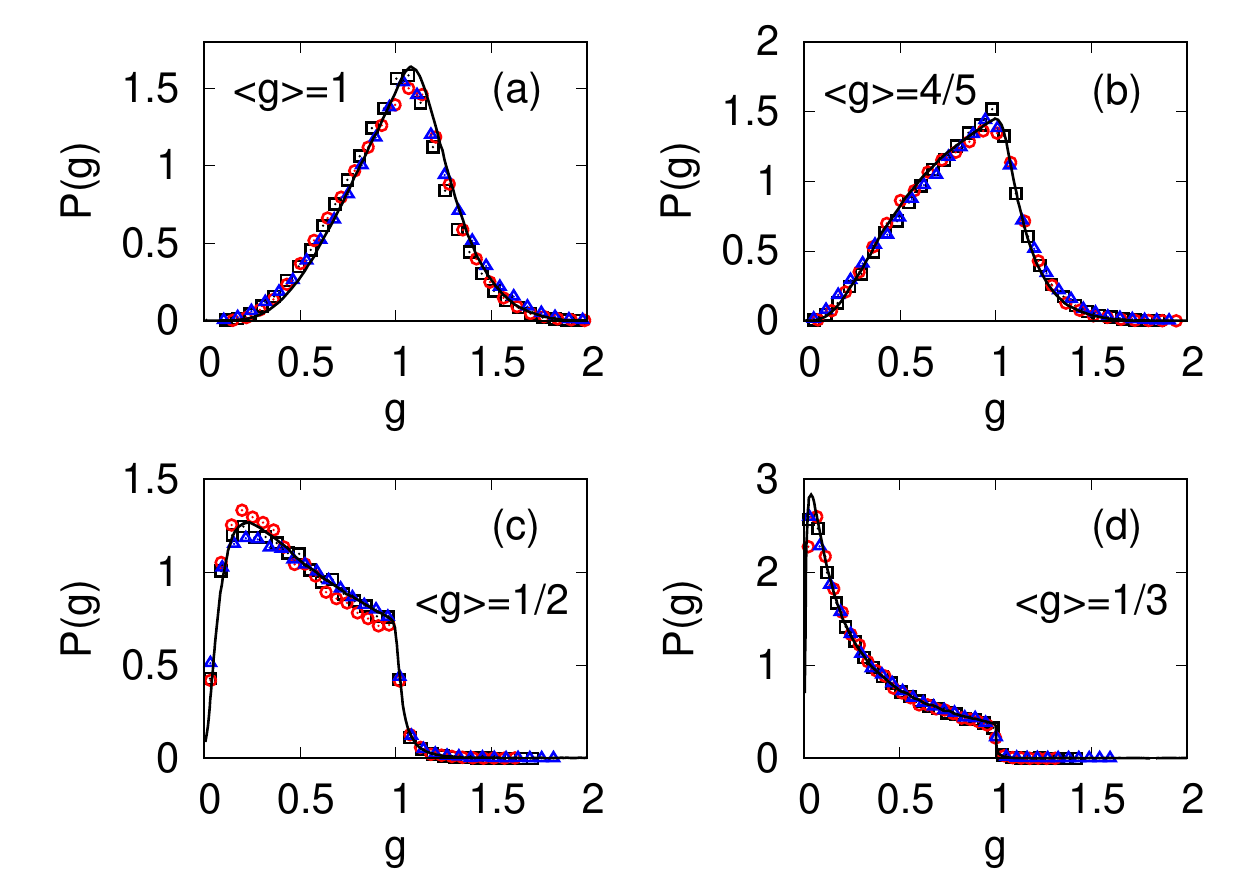}
\caption{Conductance distributions $P(g)$ in a quasi-1D system $L \times L \times L_z$ at $E = -1.35$ in the crossover regime for (a) $\langle g \rangle = 1$, (b) $\langle g \rangle = 4/5$, (c) $\langle g \rangle = 1/2$, and (d) $\langle g \rangle = 1/3$.
Here $\phi = \phi_0/3$, $t_z = 0.1$.
Three different systems were used: ($\Box$) $W = 1$, $L = 21$, ($\bigcirc$) $W = 10$, $L = 51$,  and ($\bigtriangleup$) $W = 150$, $L = 30$, corresponding to the surface disorder strength in the weak, intermediate, and strong disorder regimes.
We choose the length $L_z$ such that the resulting $\langle g \rangle$ is closest to the corresponding $\langle g \rangle$ in each plot.
The small deviation in (c) for ($\bigcirc$) $W = 10$, $L = 51$ from other curves can be attributed to its smaller $\langle g \rangle = 0.488$ compared with $\langle g \rangle = 0.50$ for other curves.
To construct each histogram, $5 \times 10^4$ disorder realizations were used.
Solid lines are the Monte Carlo solutions of the DMPK equation of a weakly disordered quasi-1D wire in the unitary class, taken from Ref.~\cite{Froufe02}.
}
\label{fig:Pg_crossover}
\end{figure}

The conductance distribution is of particular interest in the crossover regime, where $\langle g \rangle \sim 1$~\cite{Plerou98,Garcia01,Froufe02}.
Figure~\ref{fig:Pg_crossover} represents the evolution of $P(g)$ in a quasi-1D system $L \times L \times L_z$ at $E = -1.35$ with $\phi = \phi_0/3$, $t_z = 0.1$ in the crossover regime.
We choose three sets of parameters of disorder strengths and transverse system sizes, ranging from weak to strong disorder strength.
For all cases, the agreements between the three distributions are excellent.
This validates the single-parameter scaling hypothesis in this surface-disordered system.
The conductance distribution depends only on the average conductance, independent of the details of the system.
As the average conductance $\langle g \rangle$ decreases, $P(g)$ gradually deviates from the Gaussian distribution in the metallic regime.
For $\langle g \rangle = 4/5$, only the $g > 1$ part can be approximated by the Gaussian function.
At $\langle g \rangle = 1/2$, the distribution becomes highly asymmetric and there is a drastic change near $g = 1$.
Finally, for $\langle g \rangle = 1/3$, the distribution develops a huge peak in the small $g$ region, driving the system towards the insulating regime.
The peculiar forms of the conductance distributions in the crossover regime have also been observed in other systems~\cite{Plerou98,Garcia01,Froufe02,Marko02,Froufe07,Somoza09,Qiao10,Lopez18}.
Remarkably, we find that our results coincide well with the results obtained from the DMPK equation of a weakly disordered quasi-1D wire in the unitary class~\cite{Gopar02,Froufe02}, which are indicated as solid lines in Fig.~\ref{fig:Pg_crossover}.
Therefore, our results suggest that the conductance distributions of the chiral surface states in the quasi-1D regime can still be described by the DMPK equation, even in the presence of intermediate and strong surface disorder.
We note that in the strong surface disorder regime, the strong disorder on the surface layer can be effectively mapped to the weak disorder on the first inward layer, as we discussed in Sec.~\ref{subsec:Localization_length}.
Thus, the truly unexpected finding is that the description of the DMPK equation still holds in the intermediate disorder regime, where the localization length is much shorter than the transverse size of the sample. 

Finally, it is worth noting that different from ordinary surface-disordered wires~\cite{Feilhauer11}, the bulk of our system is insulating.
Thus the unique physics in surface-disordered systems, such as L\'evy flights~\cite{Leadbeater98} and the coexistence of different transport regimes~\cite{Sanchez98}, does not occur in our system.

%%%%%%%%%%%%%%%%%%%%%%%%%%%%%%%%%%%%%%%%%%%%%%%%%%%%%%%%%%%%%%%%%%%%%%%%%%
%%%%%%%%%%%%%%%%%%%%%%%%%%
\subsection{Effect of weak bulk disorder}
\label{subsec:Effect_of_bulk_disorder}
%%%%%%%%%%%%%%%%%%%%%%%%%%%%%%%%%%%%%%%%%%%%%%%%%%%%%%%%%%%%%%%%%%%%%%%%%%
%%%%%%%%%%%%%%%%%%%%%%%%%%

Up to now, we have discussed the effect of surface disorder on the surface states by assuming that the bulk is clean.
However, in a realistic sample, the bulk is often weakly disordered.
In this section, we discuss the effect of bulk disorder on the surface-disordered system by introducing weak disorder in the bulk with disorder strength $W_\mathrm{bulk} = 1$.

\begin{figure}
\centering
\includegraphics[width=\columnwidth]{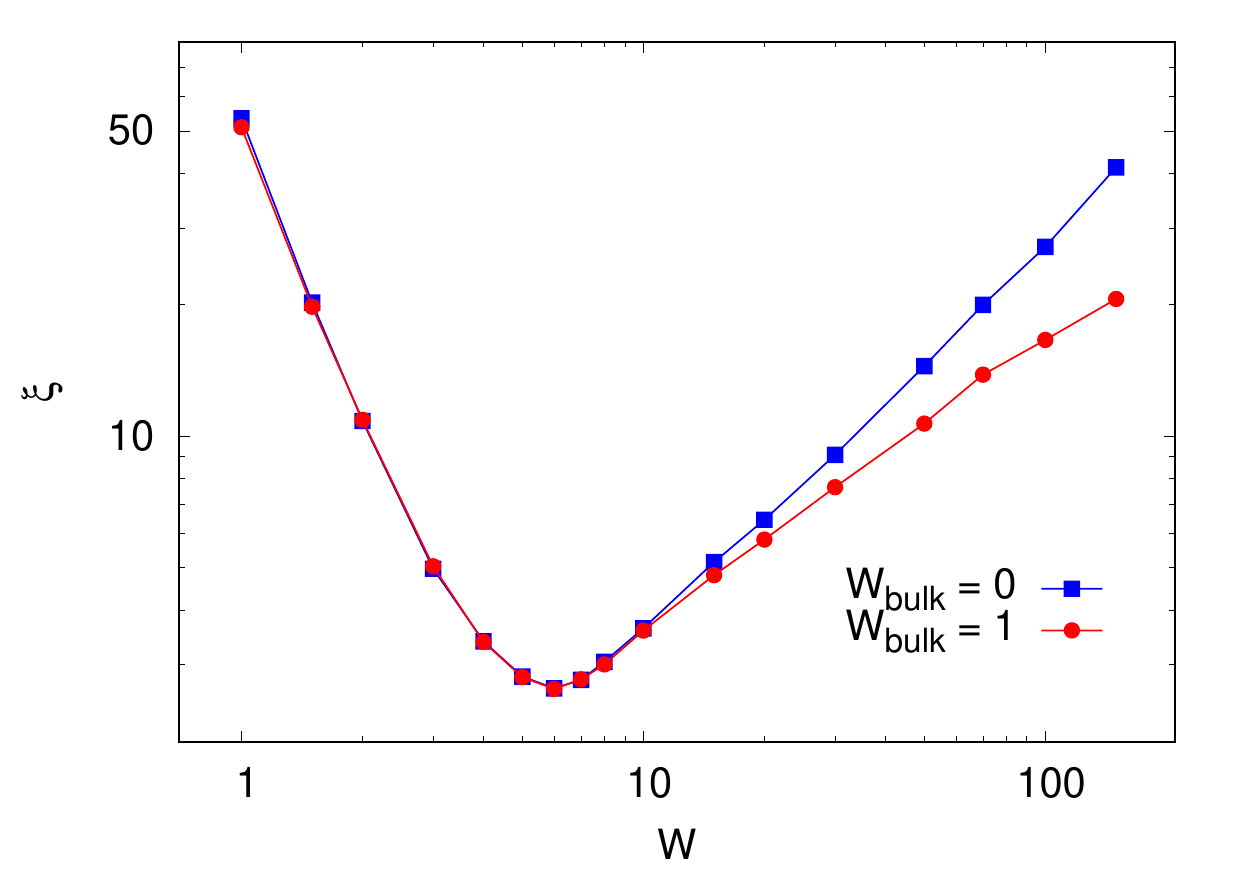}
\caption{Log-log plot of the localization length $\xi$ as a function of the surface disorder strength $W$ in a quasi-1D system $L \times L \times L_z$ at $E = -1.35$ with and without weak bulk disorder $W_\mathrm{bulk} = 1$.
Here $\phi = \phi_0/3$, $t_z = 0.1$, and $L = 21$.
In the presence of bulk disorder, the localization length almost does not change in the weak surface disorder regime and decreases significantly in the strong surface disorder regime.}
\label{fig:kesi_vs_W_with_bulk_disorder}
\end{figure}

We first discuss the effect of bulk disorder on the localization length of the surface states in the $z$ direction.
In Fig.~\ref{fig:kesi_vs_W_with_bulk_disorder}, we show the localization length as a function of the surface disorder strength $W$ in a quasi-1D system at $E = -1.35$ in the absence and presence of bulk disorder.
In the weak surface disorder regime, the weak disorder in the bulk almost does not alter the localization length.
Since the surface states mainly locate at the outermost layer in this regime, the disorder in the bulk has little influence on them.
On the other hand, the localization length decreases significantly in the strong surface disorder regime, as the surface states are now pushed into the bulk and mainly locate at the first inward layer.
In the absence of bulk disorder, one expects the localization length to be infinitely large in the limit of infinite surface disorder strength.
In this limit, the outermost layer is fully decoupled from the rest of the system, and the latter becomes disorder-free.
With bulk disorder, the localization length cannot increase unboundedly and is expected to converge to the value of a bulk-disordered system in the infinite surface disorder limit.

\begin{figure}
\centering
\includegraphics[width=\columnwidth]{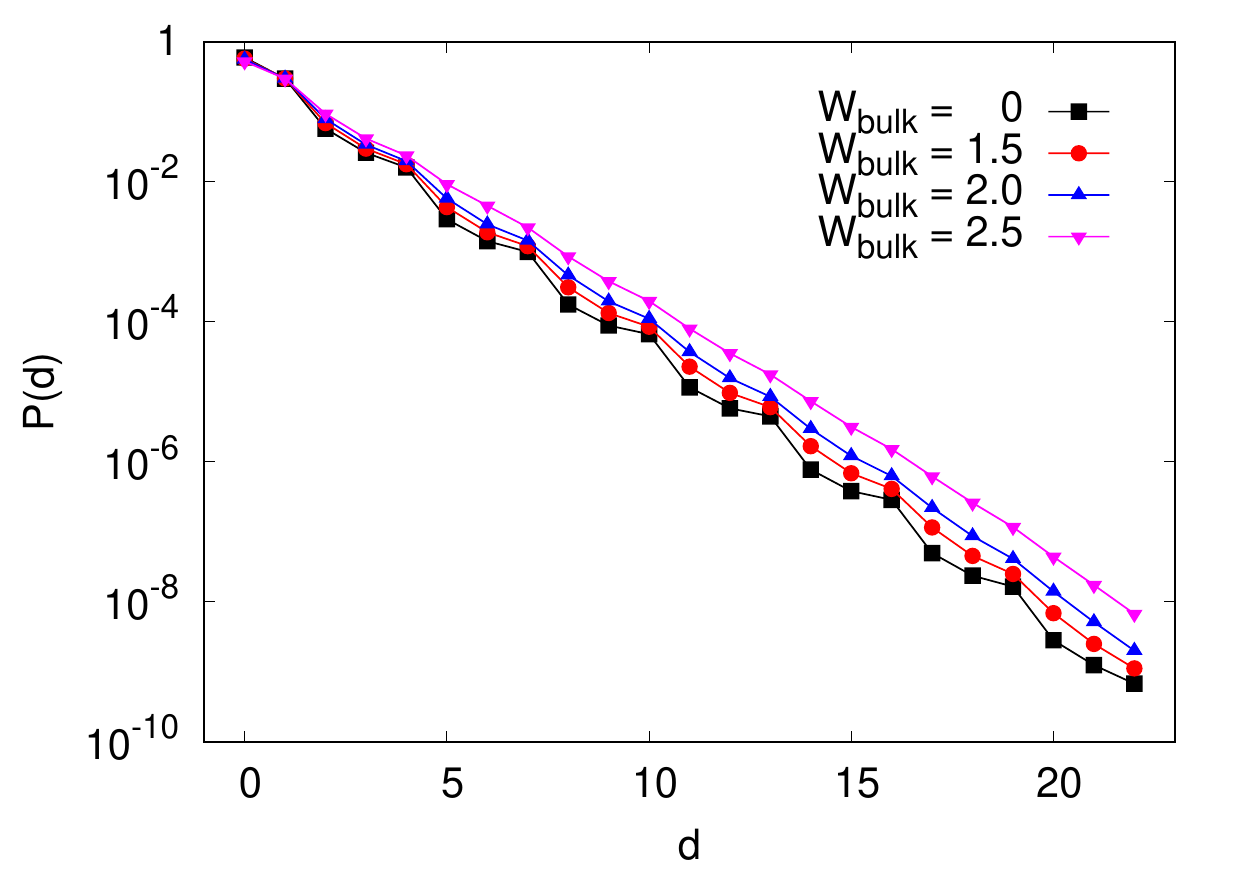}
\caption{Semi-log plot of the probability $P(d)=\int d^3 \vec{x}|\psi(\vec{x})|^2 \delta[d-d(\vec{x})]$ as a function of distance $d$ from the surface in a $48 \times 48 \times 48$ cubic lattice at $E = -1.35$ for bulk disorder strengths $W_\mathrm{bulk} = 0$, 1.5, 2, and 2.5.
Here $\phi = \phi_0/3$, $t_z = 0.1$, and the surface disorder strength $W = 6$.
We apply OBCs in the $x$ and $y$ directions and a PBC in the $z$ direction. 
The average is taken over $10^3$ disorder realizations.
As the bulk disorder strength increases, the penetration length of the surface states becomes larger.}
\label{fig:Pd_vs_w}
\end{figure}

Another effect of bulk disorder is to decrease the bulk band gap~\cite{Wang99}, hence increasing the penetration length of the surface states, which decay exponentially into the bulk.
We illustrate this in Fig.~\ref{fig:Pd_vs_w} by plotting the distributions of the surface states in a $48 \times 48 \times 48$ cubic lattice for different bulk disorder strengths $W_\mathrm{bulk}$ with the surface disorder strength $W = 6$.
Clearly, as the bulk disorder strength increases, the penetration length of the surface states becomes larger.

For conductance distributions, the presence of weak bulk disorder $W_\mathrm{bulk} = 1$ does not modify the results in Sec.~\ref{subsec:Conductance_distributions}.
We have numerically checked that $P(g)$ is still the only function of the average conductance and coincides with the DMPK results for a weakly disordered quasi-1D wire in the unitary class.

%%%%%%%%%%%%%%%%%%%%%%%%%%%%%%%%%%%%%%%%%%%%%%%%%%%%%%%%%%%%%%%%%%%%%%%%%%
%%%%%%%%%%%%%%%%%%%%%%%%%%
\section{Summary}
\label{sec:conclusions}
%%%%%%%%%%%%%%%%%%%%%%%%%%%%%%%%%%%%%%%%%%%%%%%%%%%%%%%%%%%%%%%%%%%%%%%%%%
%%%%%%%%%%%%%%%%%%%%%%%%%%

To summarize, we have investigated the effect of surface disorder on the chiral surface states of a 3D quantum Hall system.
We find that in the weak disorder regime, the localization length in the $z$ direction decreases with the surface disorder strength as expected.
However, after a critical disorder strength, which is of the order of the in-plane hopping strength, the localization length increases anomalously.
As the surface disorder strength increases, the main weight of the surface state gradually moves from the outmost layer to the first inward layer.
We explain the anomalous increase of the localization length by an effective model, which maps the strong disorder on the surface layer to the weak disorder on the first inward layer.

We also investigate the effect of surface disorder on the conductance distributions $P(g)$ of the chiral surface states in the quasi-1D regime for various surface disorder strengths.
We find that in the quasi-1D regime, the conductance distributions of the surface states can be well described by the DMPK equation of a weakly disordered quasi-1D wire, even in the presence of intermediate and strong surface disorder.
$P(g)$ is fully determined by the average conductance, independent of the surface disorder strength and the size of the system, in agreement with the single-parameter scaling hypothesis.
We have checked that the existence of weak disorder in the bulk does not modify the above physical picture.
The main effect of bulk disorder is to decrease the bulk band gap~\cite{Wang99}, hence increasing the penetration length of the surface states.
It also decreases the localization length of the surface states in the strong surface disorder regime.

Despite recent experimental developments in the 3D QHE, for real materials, the evidence of the 2D chiral surface states has only been provided in Ref.~\cite{Liu19}, partly due to the difficulty of device fabrication in the vertical transport measurement~\cite{Tang19}.
In the future, it would be interesting to further examine the properties of the surface states in experiments, especially the three transport regimes in mesoscopic samples.
Our work demonstrates that surface disorder can be an effective way to control the behavior of the surface states in the $z$ direction.
Importantly, the surface states can be pushed into the bulk and protected by the disordered surface layer in the strong surface disorder regime.
Since surface disorder can be easily manipulated by adatom deposition, ion sputtering, and air exposure, and conductance can be directly measured in experiments, we expect that our results can be verified by experiments in the future.

%%%%%%%%%%%%%%%%%%%%%%%%%%%%%%%%%%%%%%%%%%%%%%%%%%%%%%%%%%%%%%%%%%%%%%%%%%%%%%%%
\section{Acknowledgments}
\label{sec:acknowledgements}
%%%%%%%%%%%%%%%%%%%%%%%%%%%%%%%%%%%%%%%%%%%%%%%%%%%%%%%%%%%%%%%%%%%%%%%%%%%%%%%%
The work at Zhejiang University was supported by the National Natural Science Foundation of China through Grant No. 11674282 and the Strategic Priority Research Program of Chinese Academy of Sciences through Grant No. XDB28000000. K.Y.'s work was supported by DOE Grant No. DE-SC0002140, and performed at the National High Magnetic Field Laboratory, which is supported by National Science Foundation Cooperative Agreement No. DMR-1644779, and the State of Florida.

\bibliography{reference}
\end{document}